\documentclass[letterpaper]{article} 
\usepackage{aaai25}  
\usepackage{times}  
\usepackage{helvet}  
\usepackage{courier}  
\usepackage[hyphens]{url}  
\usepackage{graphicx} 
\usepackage{natbib}  
\usepackage{caption} 
\usepackage{algorithm}
\usepackage{algorithmic}

\usepackage{newfloat}
\usepackage{listings}
\usepackage{amsfonts}
\usepackage{booktabs}
\usepackage{comment}
\usepackage{amsmath}

\DeclareCaptionStyle{ruled}{labelfont=normalfont,labelsep=colon,strut=off} 
\floatstyle{ruled}
\newfloat{listing}{tb}{lst}{}
\floatname{listing}{Listing}
\usepackage{xcolor}

\title{SAFE: Self-Supervised Anomaly Detection Framework for Intrusion Detection}
\author{
    Elvin Li\equalcontrib,
    Zhengli Shang\equalcontrib,
    Onat Gungor,
    Tajana Rosing
}

\affiliations{
    University of California, San Diego\\

    9500 Gilman Dr, La Jolla, CA 92093 USA\\
    \{ell009, z4shang, ogungor, tajana\}@ucsd.edu
}

\usepackage{bibentry}
\usepackage{caption}

\begin{document}

\maketitle
\begingroup
\renewcommand\thefootnote{} 
\endgroup
\begin{abstract}
The proliferation of IoT devices has significantly increased network vulnerabilities, creating an urgent need for effective Intrusion Detection Systems (IDS). Machine Learning-based IDS (ML-IDS) offer advanced detection capabilities but rely on labeled attack data, which limits their ability to identify unknown threats. Self-Supervised Learning (SSL) presents a promising solution by using only normal data to detect patterns and anomalies. This paper introduces SAFE, a novel framework that transforms tabular network intrusion data into an image-like format, enabling Masked Autoencoders (MAEs) to learn robust representations of network behavior. The features extracted by the MAEs are then incorporated into a lightweight novelty detector, enhancing the effectiveness of anomaly detection. Experimental results demonstrate that SAFE outperforms the state-of-the-art anomaly detection method, Scale Learning-based Deep Anomaly Detection method (SLAD), by up to 26.2\% and surpasses the state-of-the-art SSL-based network intrusion detection approach, Anomal-E, by up to 23.5\% in F1-score.
 
\end{abstract}

\section{Introduction}
As computer networks are increasingly used in today's world, this trend accelerates the growth of network intrusions and threatens the reliability, availability, and stability of network services, particularly in the context of the Internet of Things (IoT) \cite{zarpelao2017survey}. To ensure the security of IoT systems, Intrusion Detection Systems (IDS) are developed to monitor network traffic, identify anomalies, and differentiate between normal activities and malicious threats \cite{gungor2024rigorous}. However, as cyber threats become more sophisticated, traditional (signature-based) IDS face limitations in identifying complex and evolving attack patterns. To address these challenges, machine learning-based IDS (ML-IDS) have emerged. ML-IDS enhance detection accuracy through their advanced capability to analyze vast amounts of network data \cite{liu2019machine}. 

\begin{figure}[t]
    \centering
    \captionsetup{justification=centering}
    \includegraphics[width=\linewidth]{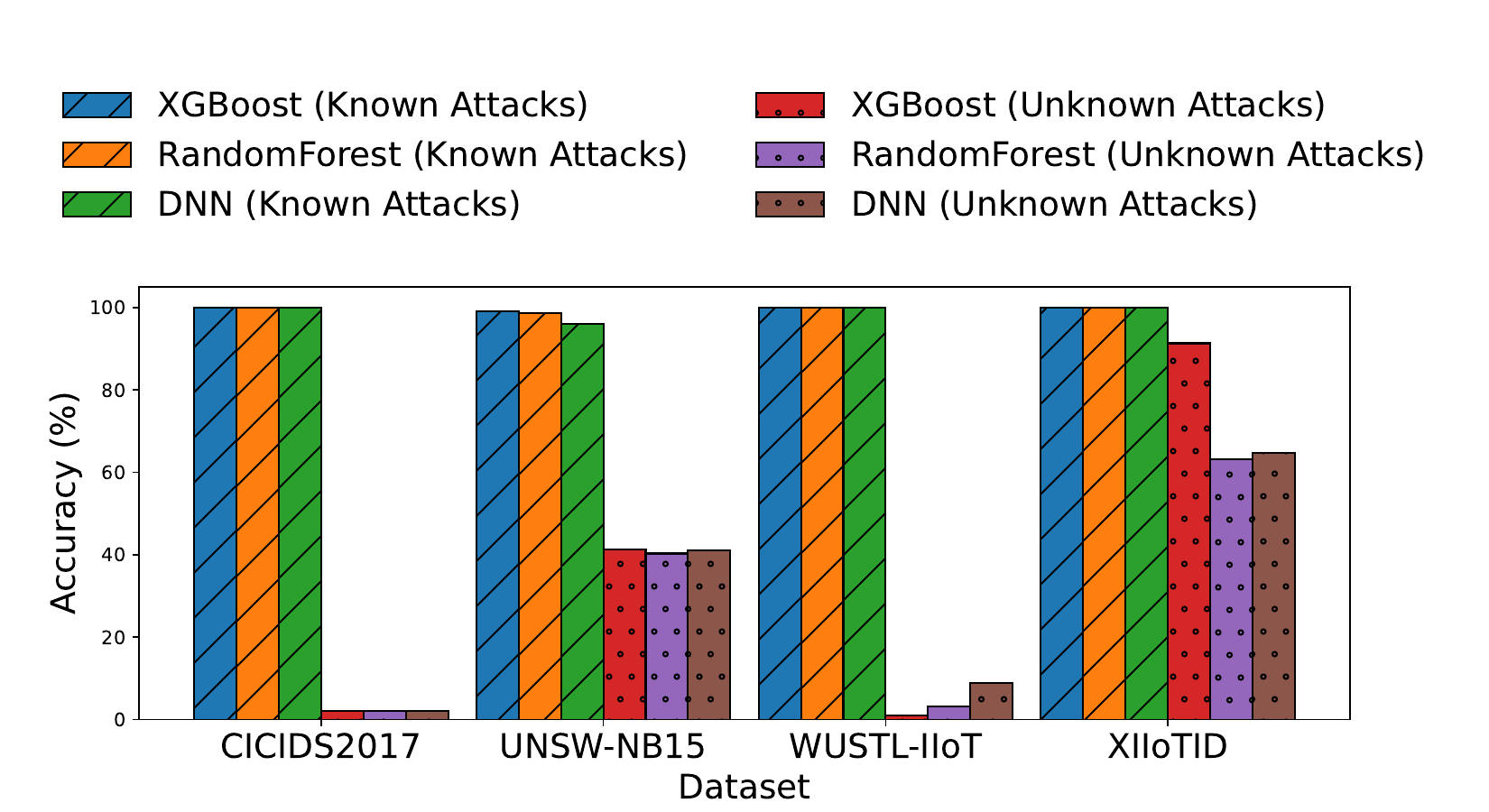}
    \caption{State-of-the-art supervised ML intrusion detection performance on known and unknown attacks}
    \label{fig:unknown_attacks}
\end{figure}

Supervised ML-based IDS rely on labeled attack data to identify network anomalies. However, their dependence on labeled data exposes a critical limitation. While supervised learning models are effective against known threats, they are unable to detect zero-day attacks, which exploit previously unknown vulnerabilities \cite{caville2022anomal}. Figure~\ref{fig:unknown_attacks} illustrates the limitations of supervised ML-IDS methods in handling zero-day attacks, revealing a substantial decline in performance when faced with previously unseen threats. On the other hand, unsupervised learning can learn from unlabeled data, which is abundant and inexpensive to obtain. However, it struggles to adapt to unfamiliar data without labeled guidance, often overfitting to training patterns and becoming less robust to distribution shifts \cite{almaraz2023enhancing}. This limitation is problematic in today's constantly shifting threat environment, where novel attacks frequently emerge \cite{karn2021learning}.

Self-supervised learning (SSL) is proposed to address the challenge of distinguishing unknown attacks and improving generalization to unseen data, offering a powerful approach for training models, particularly in fields where labeled data is scarce or difficult to obtain \cite{nguyen2023ts}. SSL can learn useful representations from unlabeled data through pre-training, where the model generates labels based on inherent patterns within the data itself. This approach enables the model to develop strong feature representations that can generalize to various tasks. SSL's ability to effectively handle unknown or evolving threats stems from its capacity to learn directly from the data, enhancing its adaptability and robustness \cite{wang2023robust}. One compelling SSL approach is the use of masked autoencoders (MAEs). MAEs have shown exceptional performance in image classification by learning rich representations \cite{he2022masked}. By masking parts of an image and training the model to reconstruct them, this approach enables the model to capture underlying patterns and spatial structures. In the context of network flow, MAEs offer similar benefits. By masking portions of network traffic data and training the model to predict the masked parts, the model learns critical patterns of normal network behavior and subtle anomalies, thereby enhancing its ability to detect emerging threats. This makes MAEs a powerful tool for identifying complex intrusions without relying on labeled attack data.

\begin{figure*}[t]
    \centering
    \includegraphics[width=.95\linewidth]{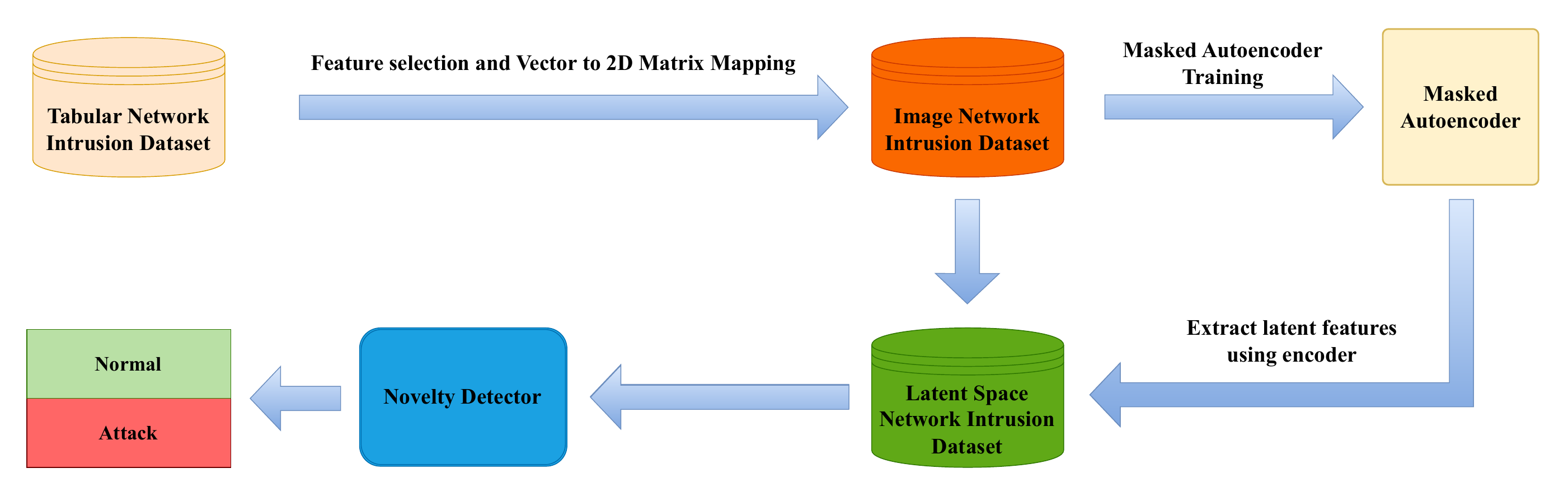}
    \captionsetup{justification=centering}
    \caption{The proposed framework (SAFE) overview. Our methodology begins with feature selection and mapping each vector data point to an image matrix. This mapped data is subsequently used to train an MAE. The encoder head of the MAE extracts latent features, which are then provided to the novelty detector, used to distinguish between attack and normal instances.}
    \label{fig:high-level}
\end{figure*}

In this paper, we propose SAFE (Figure \ref{fig:high-level}), a self-supervised anomaly detection framework for intrusion detection. The proposed framework begins by applying Principal Component Analysis (PCA) for feature selection, enhancing the data for subsequent processing. The selected features are then transformed into an image-like format by reshaping them into a structured grid, similar to the pixel arrangement in an image, thereby optimizing the data for compatibility with a masked autoencoder (MAE). 
The MAE is then trained on the transformed data to learn meaningful patterns that represent normal network behavior and help identify potential anomalies. Finally, the features extracted by the MAE are fed into a lightweight novelty detection model, Local Outlier Factor (LOF) \cite{cheng2019outlier}, to identify anomalies. This layered approach integrates robust feature selection, effective representation learning, and efficient novelty detection, enhancing the system’s ability to detect and adapt to diverse network intrusions. Our experimental results on recent intrusion detection datasets demonstrate that SAFE outperforms the state-of-the-art anomaly detection algorithm, Scale Learning-based Deep Anomaly Detection method (SLAD) \cite{xu2023fascinating}, and the state-of-the-art self-supervised intrusion detection method, Anomal-E \cite{caville2022anomal}. SAFE increases the F1-score by up to 26.2\% compared to SLAD and up to 23.5\% compared to Anomal-E (on average 17.23\% and 16.03\%, respectively).

\section{Related Works}
The growing use of IoT devices has revealed the limitations of traditional Intrusion Detection Systems (IDS), which often cannot effectively handle the diverse traffic patterns of IoT networks \cite{khan2022deep}. This has driven interest in ML-based IDS, which utilize data-driven methods to improve the detection of network anomalies. Among these ML-based approaches, supervised learning is effective for known threats but fails to detect zero-day attacks due to its reliance on patterns from labeled data. Unsupervised learning, while not relying on labeled data, still needs further investigation to enhance its generalization capabilities in network intrusion detection \cite{verkerken2022towards}. To address this gap, Self-Supervised Learning (SSL) has emerged as a promising alternative, combining the advantages of both supervised and unsupervised learning methodologies. 

Several studies focus on using contrastive learning for intrusion detection. Almaraz et al. transform IDS datasets into synthetic grayscale images, which are then augmented using techniques such as cropping, and adding noise \cite{almaraz2023enhancing}. They then pre-train using the Momentum Contrast (MoCo v2) technique and fine-tune the model for tasks such as attack detection and protocol classification. Yang et al. process untagged packet data into vectorized representations and create masked matrices to enhance pre-training \cite{yang2023malicious}. Yue et al. employ an encoder-projector-predictor architecture, mapping feature outputs onto a unit sphere to enhance classification accuracy \cite{yue2022contrastive}. 
Another promising direction is integrating graph-based algorithms for intrusion detection. Nguyen et al. propose Traffic-aware Self-supervised Learning (TS-IDS) to address the unique characteristics of IoT traffic by representing network data as a graph \cite{nguyen2023ts}. TS-IDS uses a Graph Neural Network (GNN) enriched by auxiliary property-based SSL to learn IoT-specific communication patterns. Caville et al. present the Anomal-E framework, which offers a more generalized graph-based approach for broader network intrusion detection \cite{caville2022anomal}. Anomal-E structures network flows as graphs, with nodes representing entities and edges denoting communication flows, and applies contrastive learning to maximize mutual information across local and global graph representations to distinguish between normal and anomalous patterns. 

Recent studies have explored the use of novelty detection methods to effectively distinguish between normal and abnormal network traffic. Local Outlier Factor (LOF), when applied in a network intrusion context, measures the local density of network flow data points and identifies anomalies based on deviations from expected patterns, making it effective for detecting isolated attacks or rare traffic behaviors \cite{alghushairy2020review}. Isolation Forest (IF) enhances anomaly detection in network intrusion systems by recursively partitioning traffic data, efficiently isolating anomalous patterns in large, high-dimensional network datasets \cite{al2021isolation}. Deep Isolation Forest (DIF) extends this approach by using SSL-pretrained representations of network traffic to map flows into high-dimensional spaces, improving its ability to isolate complex and subtle intrusions \cite{xu2023deep}. However, these methods struggle with the complex, dynamic nature of real-time network traffic and face scalability challenges.  

Although these studies highlight the scope of SSL-based network intrusion detection, many rely on computationally intensive models like transformers, resulting in resource overhead and reduced scalability for real-time detection. Additionally, pre-training methods using ResNet and MoCo, while effective in image-based contexts, may struggle to capture the patterns inherent in network traffic. GNN-based IDS frameworks face challenges in terms of efficiency and scalability. Constructing graph representations from raw network data is time-consuming, and GNN models further extend processing time. In contrast, by transforming tabular network flow data into an image format that maps correlated features to meaningful spatial representations, and subsequently employing a masked autoencoder within a lightweight framework, our approach is better aligned with the unique structure and demands of network datasets. 

\section{Proposed Framework}

Figure \ref{fig:high-level} presents SAFE, our self-supervised anomaly detection framework, which is composed of four main modules: feature selection, vector-to-image matrix feature mapping, masked autoencoder (MAE) training, and MAE feature extraction-based novelty detection. Motivated by the success of MAEs within the vision domain from recent years \cite{han2024efficient,he2022masked}, our core idea is to transform the original vector features into a suitable matrix, where a geometric relationship exists between each element. This is analogous to the properties of images. Since network intrusion data is inherently tabular, a method is needed to transform it into an image-like format to leverage the MAE vision capabilities. After this transformation, we train an MAE and extract features from its encoder. These features are then used to train a novelty detector, which distinguishes between attack and normal samples. Given a new sample from the test data, we use the trained MAE and novelty detector during inference to determine whether it is normal or an attack. Table \ref{tab:variable-references} introduces variable notations used in this section. \par 

\begin{table}[t]
    \centering
    \small
    \caption{Variable Notations}
    \label{tab:variable-references}
    \scalebox{0.9} {
    \begin{tabular}{ll}
        \toprule
        \textbf{Variable} & \textbf{Description} \\
        \midrule
        \textbf{$\mathcal{D} = \{(x_i,y_i) \: | \: x_i \in \mathbb{R}^d\}_{i=1}^n$}
        & Original Intrusion Dataset \\
        \midrule
        \textbf{$\mathcal{D}_F = \{(x_{fi},y_i) \: | \: x_{fi} \in \mathbb{R}^k, k \leq d\}_{i=1}^n$}
        & Feature Filtered Dataset \\
        \midrule
        \textbf{$\mathcal{D}_I = \{(x_i',y_i) \: | \: x_i' \in \mathbb{R}^{d' \times d'}\}_{i=1}^n$} 
        & Image Intrusion Dataset  \\
        \midrule
        \textbf{$\mathcal{D}_L = \{(x_i'',y_i) \: | \: x_i'' \in \mathbb{R}^{d''}\}_{i=1}^n$} 
        & Latent Space Dataset  \\
        \midrule
        \textbf{$0$} 
        & Label for Normal Data  \\
        \midrule
        \textbf{$1$} 
        & Label for Attack Data  \\
        \bottomrule
    \end{tabular}}
\end{table}

\subsection{Module 1: Feature Selection} Selecting an effective feature selection algorithm is crucial for reducing dimensionality, thereby facilitating the pre-training of an MAE with reduced computational overhead. Moreover, given that the latent space generated by the MAE inherently possesses fewer dimensions than the original feature space, it is imperative to ensure that only the most salient features contribute to its construction. We utilize a key property of principal component analysis (PCA) to perform dimensionality reduction \cite{song2010feature}. In addition to its low computational cost, we specifically choose PCA because it is an unsupervised technique, eliminating the need for prior knowledge of network intrusions. Consider $\mathcal{D}_{train} = \{(x_{i_{train}},y_{i_{train}}) \: | \: x_i \in \mathbb{R}^{d_{train}}, y_{i_{train}} = 0\}_{i=1}^{n_{train}}$ to be the training set of our original dataset. 
Since every principal component $PC_i$ is a linear combination of the original features $x_{i_{train}}$, denote it by $PC_j = a_{j,1}x_{1} + a_{j,2}x_{2} + a_{j,3}x_{3} + ... + a_{j,d_{train}}x_{j,d_{train}}$ where $j$ indicates the $j$th principal component, and $a_{j,i}, i \in \{1,...,d_{train}\}$ are the loadings of each feature for the respective $j$th principal component. We then calculate the ranking of each $x_{i_{train}}$ through $\mathcal{R}(x_{i_{train}}) = \sum_{j=1}^M |a_{j,i}|$, where $M$ is the total number of principal components. To select $M$, we pick the least number of principal components which cumulatively account for an explained variance ratio, the latter of which is set to $95\%$. Applying $\mathcal{R}(x_{i_{train}})$ to each feature, we obtain a numerical set that can be sorted in descending order to represent the importance of each feature. Select an arbitrary $k \in \mathbb{N}, 1 \leq k \leq d$, and retain only the top $k$ features ranked by $\mathcal{R}(x_{i_{train}})$ to obtain $\mathcal{D}_F$.

\subsection{Module 2: Vector to Image Matrix Mapping} As discussed earlier, MAE benefits have been demonstrated in the vision domain; therefore, it is necessary to train the model on an image dataset. Since network intrusion data is tabular, we need to create a mapping for each vector datapoint into a subsequent image matrix counterpart. This means that each feature within the transformed image matrix should have a spatial relationship with neighboring features, which outlines our goal to transform from $\mathcal{D}_F$ to $\mathcal{D}_I$. \par 
We leverage DeepInsight \cite{9701572} to efficiently transform the feature vectors into feature matrices and create a map between each datapoint $(x_i, y_i)$ and their respective transformation $(x_i',y_i)$. 
To achieve this, our configuration of DeepInsight utilizes t-Stochastic Neighbor Embedding (t-SNE) \cite{van2008visualizing} to reduce each feature's vector representation from $\mathbb{R}^n$ to $\mathbb{R}^2$. This allows us to find the $\mathbb{R}^2$ coordinate locations of each specific feature, $x_j$, which we can denote as $(a_j, b_j)$. If a bijection between features and their coordinates does not exist, discretization techniques such as linear sum assignment are applied to establish a one-to-one correspondence, ensuring that the coordinates for our features remain unique. DeepInsight then applies a convex hull algorithm to generate the largest polygon that encapsulates all of the $\mathbb{R}^2$ feature representation points. Performing this, along with then reshaping the data into a $d' \times d'$ square matrix, creates a finite space that contains every feature representation whilst having the smallest area (thereby reducing the number of irrelevant, empty coordinates). Then, we iterate through our datapoints and for each $(x_{fi},y_i)$ map every feature $x_{fi,s} \in x_{fi}, s \in \{1,...,k\}$ to $(a_j,b_j)$ of the $\mathbb{R}^{d' \times d'}$ matrix in accordance to the original intensity value, normalized as an integer between $[0,255]$ (the standard numerical color value range). After applying the t-SNE based tabular image conversion (trained on $\mathcal{D}_{F_{train}} = \{(x_{fi_{train}},y_{i_{train}}) \: | \: x_i \in \mathbb{R}^{k_{train}}, y_{i_{train}} = 0\}_{i=1}^{n_{train}}$) to the entire dataset of $\mathcal{D}_F$, we obtain $\mathcal{D}_I$ (image intrusion dataset).
\par 

\subsection{Module 3: MAE Training} 
\begin{figure}[t]
    \centering
    \captionsetup{justification=centering}
    \includegraphics[width=\linewidth]{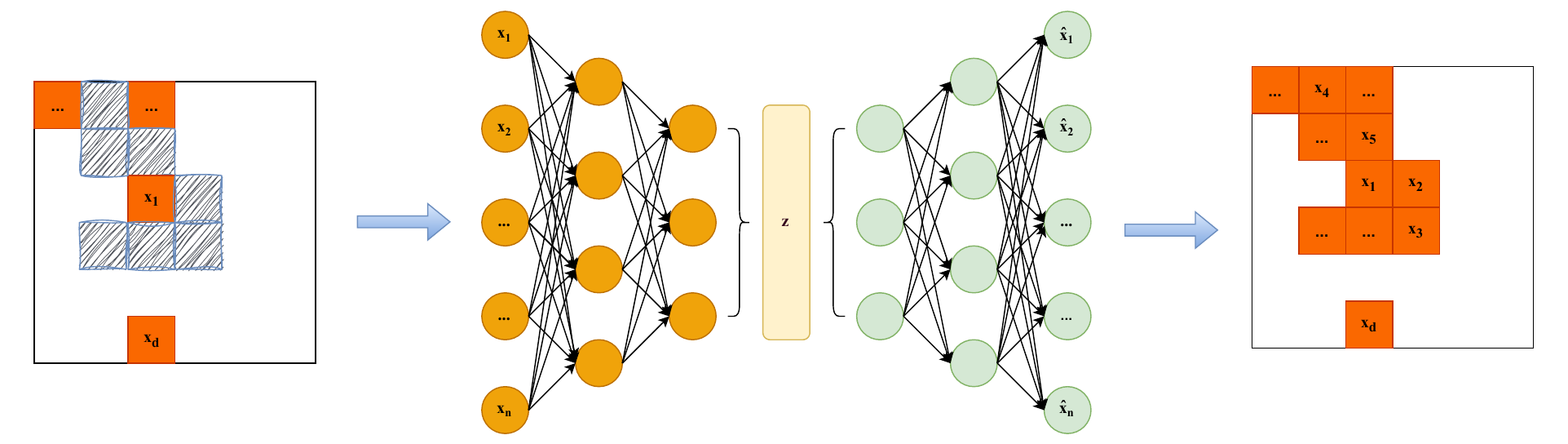} 
    \caption{Visual representation of the masking operation. The MAE is trained to learn how to reconstruct the image when a random portion of the pixels are set to value $0$.}
    \label{fig:mae-example}
\end{figure}

Given that $\mathcal{D}_I$ consists of matrices with spatial relationships between entries, it is well-suited for use as a vision dataset. Figure \ref{fig:mae-example} highlights what we aim to accomplish in this module, which is to pre-train an MAE with $\mathcal{D}_{I_{train}}=\{(x'_{i_{train}},y_{i_{train}}) \: | \: x'_{i_{train}} \in \mathbb{R}^{d' \times d'}\}_{i=1}^{n_{train}}$ to learn its latent representation. 
For our MAE architecture, we designed an encoder-decoder structure that processes inputs through convolutional layers and randomly masks training inputs according to a fixed ratio. Each training datapoint was reconstructed after the mask was applied, and the loss for the reconstructed image was then evaluated through the mean squared error (MSE) of individual masked features between the original datapoint and the reconstructed one. 
We also construct a bottleneck layer, known colloquially as the latent space, which serves as the final layer of the encoder and the input layer to the decoder.  
Then, we train our MAE on $\mathcal{D}_{I_{train}}$. Consistent with the typical behavior of autoencoders, it learns a compact and latent representation of the features which is provided to our novelty detector. 

\subsection{Module 4: Feature Extraction and Novelty Detection} 
Given the trained MAE from Module 3, let us consider its encoder head $\mathcal{H}$, which now incorporates modified weights. In this module, we explain how we use $\mathcal{H}$ to obtain $\mathcal{D}_L$ (latent space features), and then subsequently apply a novelty detector as our penultimate step to classify attacks. \par  
Since $\mathcal{H}$ outputs to a latent space, we effectively input $\mathcal{D}_I$ through the encoder, which will take every image datapoint inside and convert them into a latent vector. Combining these individual latent vector datapoints, they collectively form $\mathcal{D}_L$. We can use these new features to train a novelty detector. Specifically, consider $\mathcal{D}_{L_{train}} = \{(x''_{i_{train}},y_{i_{train}}) \: | \: x''_{i_{train}} \in \mathbb{R}^{d''}, y_{i_{train}} = 0\}_{i=1}^{n_{train}}$. We use $\mathcal{D}_{L_{train}}$ to train a novelty detector, $\mathcal{C}$, which serves as our final attack classifier. We specifically chose $\mathcal{C}$ to be Local Outlier Factor (LOF) \cite{cheng2019outlier}. This decision was based on the fact that LOF identifies anomalies by assessing local density deviations, rather than relying on global distance metrics, thereby facilitating anomaly detection through the evaluation of a data point’s density relative to its local neighborhood. In the latent space where only essential feature representations are preserved, the local structure surrounding each datapoint becomes clearer and more compact, thereby improving the efficacy of LOF. Additionally, LOF exhibits a computationally efficient runtime compared to many other anomaly detection methods, rendering it particularly well-suited for the network intrusion domain. \par
\textbf{Inference:} With our final trained novelty detector $\mathcal{C}$, we can use it to detect anomalies from incoming network flow data. During inference, we apply the results from modules 1-2 to process each new datapoint into an image format, and then use $\mathcal{H}$ to create its latent space vector representation. It is then in a suitable format for which we apply $\mathcal{C}$ to classify the new datapoint as either an attack or normal network flow. 

\section{Experiments}

\subsection{Datasets}
We evaluate SAFE using four different intrusion datasets, which are summarized in Table \ref{datasets} and described as follows: \\
\textbf{MQTTset:} MQTTset focuses on the MQTT protocol commonly used in IoT communications. It provides labeled data for evaluating security mechanisms against MQTT-specific vulnerabilities and attacks \cite{vaccari2020mqttset}. \\
\textbf{WUSTL-IIoT:} Developed by Washington University, the WUSTL-IIoT dataset includes traffic data from IIoT environments, capturing a range of operational and attack scenarios that highlight the security challenges in industrial systems \cite{zolanvari2021wustl}. \\ 
\textbf{X-IIoTID:} X-IIoTID is a benchmark dataset designed for evaluating anomaly detection in Industrial Internet of Things (IIoT) networks. It encompasses diverse traffic patterns and attacks, providing a realistic testbed for IIoT security research \cite{9504604}. \\
\textbf{Edge-IIoTset:} This dataset focuses on edge computing in IIoT, containing various attack types and normal data patterns from edge devices. It is structured to support anomaly detection and cybersecurity research specific to edge environments \cite{ferrag2022edge}. 

\begin{table}[]
\centering
\caption{Selected intrusion datasets}
\scalebox{0.85}{
\begin{tabular}{|c|c|c|c|c|}
\hline
\textbf{Dataset} & \textbf{Year} & \textbf{\begin{tabular}[c]{@{}c@{}}Number of \\ Features\end{tabular}} & \textbf{\begin{tabular}[c]{@{}c@{}}Number of \\ Attacks\end{tabular}} & \textbf{\begin{tabular}[c]{@{}c@{}}Number of \\ Samples\end{tabular}} \\ \hline
MQTTset          & 2020          & 33                                                                     & 5                                                                     & 20M                                                                   \\ \hline
WUSTL-IIoT        & 2021          & 41                                                                     & 4                                                                     & 1M                                                                    \\ \hline
X-IIoTID         & 2021          & 59                                                                     & 18                                                                    & 0.9M                                                                  \\ \hline
Edge-IIoTset    & 2022          & 61                                                                     & 14                                                                    & 2.2M                                                                  \\ \hline
\end{tabular}}
\label{datasets}
\end{table}

\subsection{Baselines} We compare our solution with three traditional novelty detection algorithms and six state-of-the-art deep anomaly detection models. We train these models on the original features and subsequently fine-tune them using Bayesian optimization \cite{wu2019hyperparameter} to enhance their performance, thereby enabling fair comparisons with SAFE. \\
\textbf{LOF: } Local Outlier Factor (LOF) calculates the local density deviation of a sample relative to its neighbors, identifying points with significantly lower density than their surroundings as anomalies \cite{cheng2019outlier}. \\
\textbf{IF: } Isolation Forest (IF) isolates anomalies by constructing random decision trees. Anomalies, being easier to isolate than normal points, typically exhibit shorter path lengths within these trees \cite{liu2008isolation}. \\ 
\textbf{PCA: } Principal Component Analysis (PCA) reconstructs the input data and measures the reconstruction error. Points with high reconstruction error indicate anomalies. We determine the threshold value by tuning the percentile on the validation set, which is then used to classify the samples as normal or attack \cite{ndiour2020out}.   \\
\textbf{Anomal-E: }Anomal-E is a framework designed to detect anomalies using graph neural networks (GNNs). It is the first network intrusion detection method to employ GNNs in a self-supervised manner \cite{caville2022anomal}. \\ 
\textbf{SLAD: }Self-Supervised Learning Anomaly Detection (SLAD) leverages self-supervised learning to build a feature representation space, then uses these representations to detect anomalies \cite{10.5555/3618408.3620019}. \\
\textbf{ICL: }Instance-conditioned learning (ICL) captures instance-conditioned representations to improve anomaly detection by minimizing class overlap, especially in datasets with intricate patterns \cite{shenkar2022anomaly}. \\
\textbf{RCA: } Representation-Constrained Autoencoder (RCA) integrates constraints into the autoencoder structure, aiming to learn representations that are effective for distinguishing normal data from anomalies \cite{ijcai2021p208}. \\
\textbf{RDP: }Representation learning for Dynamic Predictions (RDP) is designed to improve unsupervised anomaly detection by learning representations that capture temporal or structured changes within data \cite{wang2020unsupervisedrepresentationlearningpredicting}. \par

\subsection{Experimental Setup}
\begin{figure}[t]
    \centering
    \captionsetup{justification=centering}
    \includegraphics[width=1\linewidth]{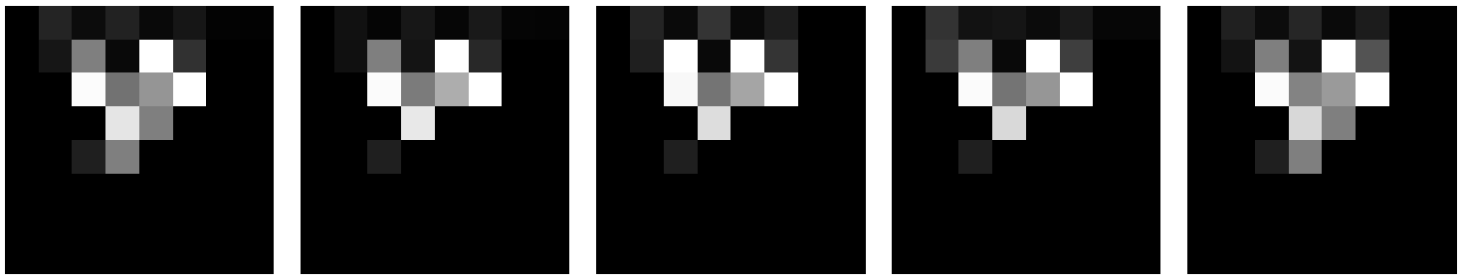}
    \includegraphics[width=1\linewidth]{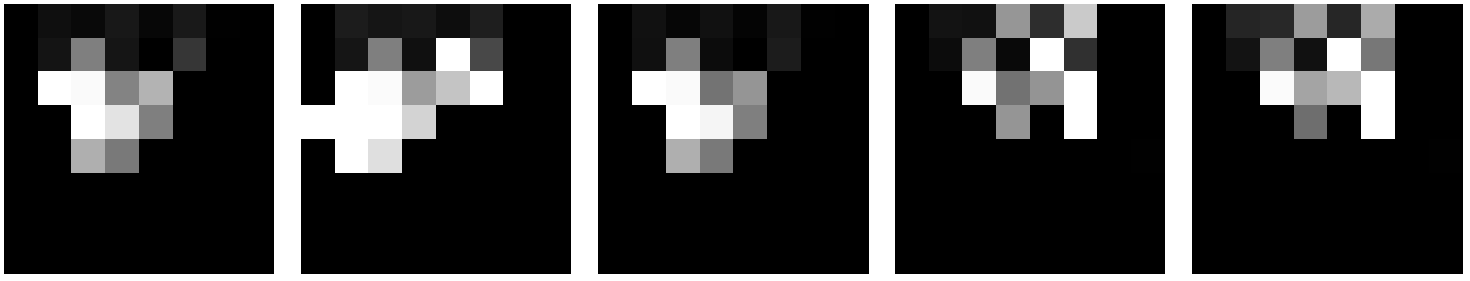}
    \caption{Selected image matrices of normal data (top) and attack data (bottom) from X-IIoTID. Even upon superficial examination, the image matrices of normal data exhibit visual homogeneity. In contrast, the attack data manifests in various forms, resulting in transformations that intuitively appear distinct.}
    \label{fig:xiiot-images}
\end{figure}

\textbf{Hardware:} We ran our experiments on a Linux virtual machine server equipped with a 32-core CPU, 64GB of RAM, and an NVIDIA RTX 2080Ti GPU. \\ 
\textbf{Data Division:} For each dataset, we split the data into 60\% training, 20\% validation, and 20\% testing sets. \\
\textbf{Image Conversion:} Using only $\mathcal{D}_{train}$ of each respective dataset, we applied Module 1 to select the top $k=31$ features. 
Then, the selected features are inputted to DeepInsight with the t-SNE, allowing us to convert them into their grayscale image representation. We chose to set our feature matrix dimension as $8 \times 8$ across all datasets. 
Figure \ref{fig:xiiot-images} illustrates the representation of X-IIoTID dataset, demonstrating the image matrix conversion for both attack and normal data. While attack data manifests in various forms, normal network flow remains homogeneous. This visual differentiation demonstrates the potential of the image transformation to aid in the detection of attacks by highlighting anomalies that deviate from the homogeneity observed in normal data. \\ 
\textbf{Training Parameters:} We trained the MAE with $20$ epochs, the Adam optimizer, and an MSE loss function for the masked portions. For each datapoint, we randomly mask $75\%$ of its entries, and then pass it through the MAE layers. For LOF training, we apply Optuna's implementation of the Tree-based Parzen Estimator for hyperparameter optimization \cite{akiba2019optunanextgenerationhyperparameteroptimization}. \\ 
\textbf{Evaluation Metrics:} We report precision, recall, and F1-score as the primary metrics for evaluation. 
\[
\text{Precision} = \frac{\text{True Positives}}{\text{True Positives} + \text{False Positives}}
\]

\[
\text{Recall} = \frac{\text{True Positives}}{\text{True Positives} + \text{False Negatives}}
\]

\[
\text{F1-score} = 2 \cdot \frac{\text{Precision} \cdot \text{Recall}}{\text{Precision} + \text{Recall}}
\]

\begin{table*}[t]
    \centering
    \captionsetup{justification=centering}
    \caption{State-of-the-art F1-score Comparison. Bold and underlined cases indicate the best and the second best, respectively. These results show that SAFE consistently outperforms other methods across the selected intrusion detection datasets.}
    \label{tab:combined-performances}
    \scalebox{0.9}{
    \begin{tabular}{lcccccc}
        \toprule
        \textbf{Model} & \textbf{X-IIoTID} & \textbf{WUSTL-IIoT} & \textbf{Edge-IIoTset} & \textbf{MQTTset} & \textbf{Average} & \textbf{Std. Dev.}\\
        \midrule
        \textbf{SAFE} & \textbf{96.41\%} & \textbf{99.40\%} & \textbf{99.95\%} & \textbf{91.38\%} & \textbf{96.79\%} & \textbf{3.92\%} \\
        \textbf{Anomal-E} & 79.44\% & 86.54\% & 80.92\% & \underline{87.53\%} & 83.61\% & \underline{4.02\%} \\
        \textbf{LOF} & 74.14\% & 96.25\% & \underline{99.94\%} & 44.69\% & 78.75\% & 25.41\% \\
        \textbf{IF} & 72.09\% & \underline{98.52\%} & 88.48\% & 79.79\% & 84.72\% & 11.38\% \\
        \textbf{PCA} & \underline{89.65\%} & 97.45\% & 97.73\% & 85.04\% & \underline{92.46\%} & 6.20\% \\
        \textbf{SLAD} & 85.71\% & 94.94\% & 79.54\% & 72.44\% & 83.16\% & 9.54\% \\
        \textbf{ICL} & 87.41\% & 93.12\% & 94.73\% & 57.48\% & 83.18\% & 17.42\% \\
        \textbf{RCA} & 81.68\% & 94.74\% & 94.46\% & 79.44\% & 87.58\% & 8.16\% \\
        \textbf{RDP} & 86.28\% & 93.07\% & 87.56\% & 71.48\% & 84.59\% & 9.23\% \\
        \bottomrule
    \end{tabular}}
\end{table*}

\begin{figure}[t]
    \centering
    \captionsetup{justification=centering}
    \includegraphics[width=1\linewidth]{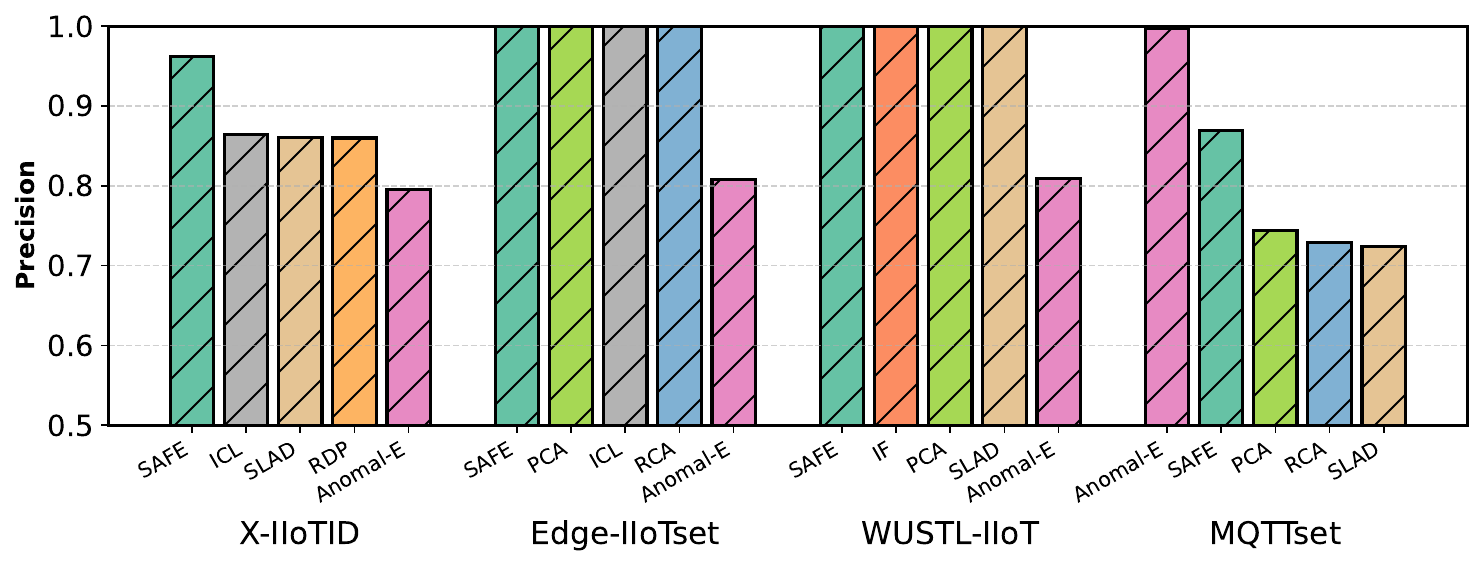}
    \includegraphics[width=1\linewidth]{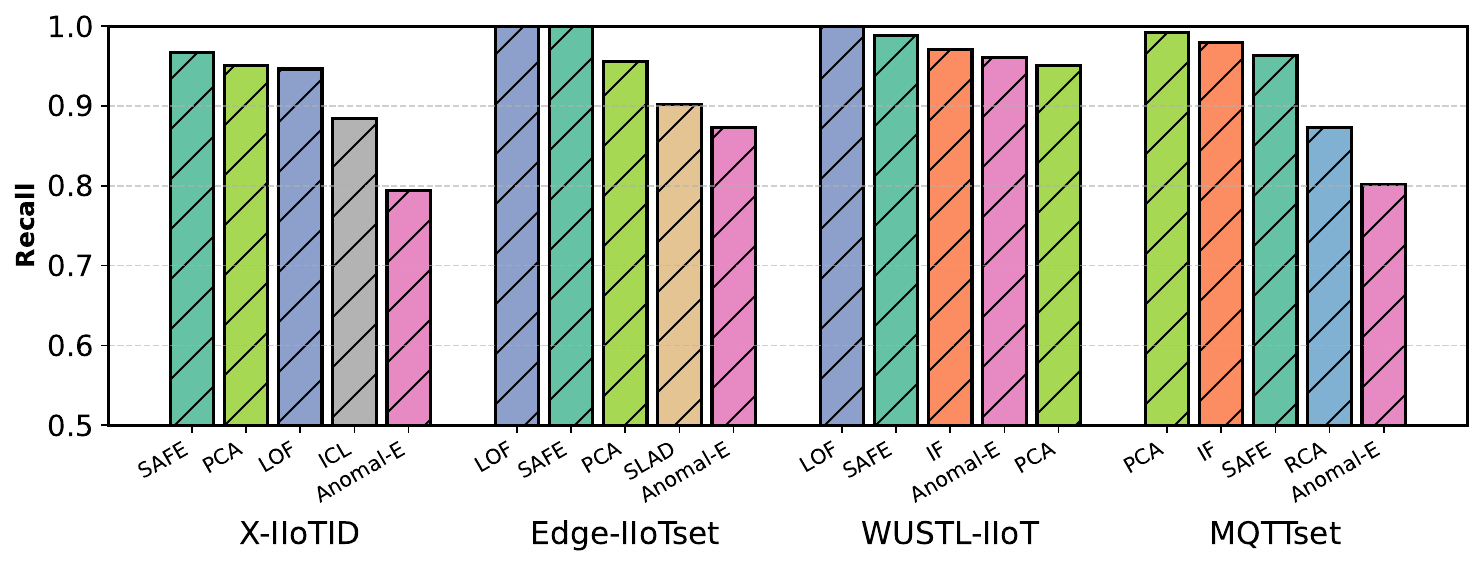}
    \caption{State-of-the-art Precision and Recall Comparison}
    \label{fig:precision-recall}
\end{figure}

\section{Results}
\label{exp-results}
\subsection{State-of-the-art Comparison}
Figure \ref{fig:precision-recall} presents a state-of-the-art comparison of precision and recall performances. To illustrate their differences, we retain the top four performers from each dataset and include Anomal-E for comparison. In zero-day attack detection, it is often more critical to minimize false negatives, as failing to detect such an attack can lead to severe security breaches. Consequently, a higher false positive rate is generally more acceptable in order to ensure that zero-day attacks are not missed. Remarkably, SAFE consistently ranks within the top three across all datasets in terms of recall, demonstrating its suitability for real-world network intrusion scenarios. Furthermore, SAFE consistently ranks within the top two for precision across all datasets, highlighting its robustness in accurately detecting intrusions. \par 
Table \ref{tab:combined-performances} presents a state-of-the-art comparison of F1-scores for each model across all datasets. Notably, a comparison of SAFE's results with those of the previously highest-performing method reveals a consistent performance improvement across all datasets: X-IIoTID (+6.76\%), Edge-IIoTset (+0.01\%), WUSTL-IIoT (+0.88\%), and MQTTset (+3.85\%). More importantly, this demonstrates that SAFE outperforms all nine selected models across the four datasets, underscoring the effectiveness of our methodology. This can be attributed to the layered feature processing mechanisms in SAFE. By systematically retaining only the most relevant features in Module 1 and enabling MAEs to process these selected features in Module 2, the MAE trained in Module 3 is able to effectively extract spatial patterns and high-quality network intrusion behaviors from the data. As a result, the features we constructed offer a more accurate representation of normal network flow compared to the original dataset features.

\begin{table}[t]
    \centering
    \captionsetup{justification=centering}
    \caption{Importance of Feature Selection (FS)}
    \small
    \label{tab:feature-selection}
    \scalebox{0.8}{
    \begin{tabular}{lcccc}
        \toprule
        \textbf{Model} & \textbf{X-IIoTID} & \textbf{WUSTL-IIoT} & \textbf{Edge-IIoTset} & \textbf{MQTTset} \\
        \midrule
        \textbf{SAFE (FS)} & \textbf{96.41\%} & \textbf{99.40\%} & \textbf{99.95\%} & \textbf{91.38\%} \\
        \textbf{SAFE (No FS)} & 94.92\% & 98.31\% & 99.82\% & 91.03\% \\
        \bottomrule
    \end{tabular}}
\end{table}

\subsection{Ablation Study} 
\hspace*{0.18cm} \textbf{Importance of Feature Selection:} Module 1 of our framework uses PCA for initial feature selection, ranking feature importance to reduce dimensionality. 
To evaluate the significance of the feature selection step, Table \ref{tab:feature-selection} presents the performance of SAFE both with and without feature selection. We observed a modest improvement in the F1-score for each dataset's novelty detector with the inclusion of feature selection: X-IIoTiD $(+1.49\%)$, WUSTL-IIoT $(+1.09\%)$, Edge-IIoTset $(+0.13\%)$, and MQTTset $(+0.30\%)$. These results underscore the significance of feature selection in our framework, highlighting its role in enhancing model performance across diverse datasets.  

\begin{table}[t]
    \centering
    \small
    \caption{SAFE Novelty Detector Selection (MQTTset)}
    \label{tab:novelty-detectors}
    \scalebox{0.85} {
    \begin{tabular}{lcccccc}
        \toprule
        \textbf{Detector} & \textbf{LOF} & \textbf{IF} & \textbf{PCA} & \textbf{SLAD} & \textbf{ICL} & \textbf{RDP} \\
        \midrule
        \textbf{F1-Score} & \textbf{91.38\%} & \underline{90.08\%} & 88.48\% & 83.24\% & 79.67\% & 84.00\% \\
        \bottomrule
    \end{tabular}}
\end{table}

\textbf{SAFE Novelty Detector Selection:} The choice of the novelty detector in Module 4 affects the prediction performance of our framework during inference. Table \ref{tab:novelty-detectors} presents the performance of the six top-performing novelty detectors on MQTTset, assuming they were selected in place of LOF. We specifically selected MQTTset for this study due to its status as the largest dataset in our collection. Furthermore, all other parameters in this study were held constant across Modules 1-3 to ensure a fair comparison. Based on our analysis, we found that LOF achieved the highest performance with an F1-score of 91.38\%, closely followed by IF, which showed a difference of 1.30\%. The other methods exhibited significantly lower performance when substituted, with ICL showing a performance difference of up to 11.71\%.  

\subsection{Overhead Analysis}

\begin{table}[t]
    \centering
    \small
    \caption{Inference Time per Sample (X-IIoTID)}
    \label{tab:overhead-analysis}
    \scalebox{0.85} {
    \begin{tabular}{lcccccc}
        \toprule
        \textbf{Detector} & \textbf{SAFE} & \textbf{Anomal-E} &\textbf{LOF} & \textbf{IF} & \textbf{PCA} & \textbf{RDP} \\
        \midrule
        \textbf{Time (ms)} & 0.1819 & 0.2436 & 0.1837 & 0.0880 & 0.0008 & 0.0155 \\
        \bottomrule
    \end{tabular}}
\end{table}

Computational overhead of ML models in IDS systems is critical because it directly impacts the system's ability to detect threats in real-time without introducing significant delays, ensuring timely responses to mitigate potential security breaches. In this context, Table \ref{tab:overhead-analysis} presents the inference time per sample overhead analysis for the top-performing methods on the X-IIoTID dataset. Despite incorporating LOF in its final module, SAFE introduces lower overhead than LOF, as it reduces data dimensionality into a compact latent representation, enabling faster classification. In this dataset, a sample corresponds to the time interval between the initial and final packets observed within a network traffic flow, with the median duration of such flows being $4.98$ milliseconds \cite{gungor2024roldef}. Relative to the median traffic flow duration, the overhead introduced by SAFE is negligible, with our method's overhead being only 3.65\%.

\section{Conclusion}
The rise of IoT devices has heightened the demand for effective network security, yet traditional IDS and ML-based IDS face critical challenges, including dependency on labeled attack data and limited adaptability to zero-day threats. Self-supervised learning (SSL) addresses these gaps by enabling models to learn patterns from unlabeled data, enhancing their capacity to identify emerging threats. We presented SAFE, a novel framework that restructures tabular network data into an image-like representation, allowing Masked Autoencoders (MAEs) to extract comprehensive insights into network behavior. This approach, combined with a lightweight anomaly detection mechanism, demonstrated superior performance in our experiments, surpassing state-of-the-art novelty detector, SLAD, by 26.15\% and outperforming state-of-the-art SSL approach, Anomal-E, by 23.52\% in F1-score. These results highlight the effectiveness of our approach in addressing the limitations of existing systems while enhancing the detection of complex and emerging threats.

\section{Acknowledgements}
This work has been funded in part by NSF, with award numbers \#1826967, \#1911095, \#2003279, \#2052809, \#2100237, \#2112167, \#2112665, and in part by PRISM and CoCoSys, centers in JUMP 2.0, an SRC program sponsored by DARPA.

\bibliography{aaai25}

\end{document}